\documentclass[12pt,fleqn]{article}
\usepackage{epsfig,rotating}

\begin{document}
\title{\bf Physics in a Mirror:\\ 
 The TRIUMF 221 MeV $pp$ Parity Violation Experiment}

\author{W.D.~Ramsay, J.~Birchall, J.~Bland, A.A.~Hamian, L.~Lee,
S.A.~Page, \\
W.T.H.~van~Oers, R.J.~Woo\\
\vspace*{0.2cm}
{\small Department of Physics, University of Manitoba, 
              Winnipeg MB, Canada R3T 2N2}\\
P.W.~Green, G.M.~Roy, G.M.~Stinson\\
\vspace*{0.2cm}
{\small Department of Physics, University of Alberta, Edmonton
AB, Canada T5G 2N5}\\
Y.~Kuznetsov, N.A.~Titov, A.N.~Zelenski \\
\vspace*{0.2cm}
{\small Institute for Nuclear Research, Russian Academy of Sciences,
117312 Moscow, Russia}\\
J.D.~Bowman, R.E.~Mischke\\
\vspace*{0.2cm}
{\small Los Alamos National Laboratory, Los Alamos NM, USA 87545}\\
C.A.~Davis, R.~Helmer, C.D.P.~Levy \\
\vspace*{0.2cm}
{\small TRIUMF, 4004 Wesbrook Mall, Vancouver BC, Canada V6T 2A3}\\
}

\date{TRI-PP-99-02}

\maketitle

\begin{abstract} 
We measure the difference in the scattering probability when an
experiment scattering longitudinally polarized 221 MeV protons from
liquid hydrogen is replaced by its mirror image. The result depends on
the interplay between the weak and strong interactions in the
interesting region near the surface of the proton. The experiment is
technically very challenging and requires elaborate precautions to
measure and correct for various sources of systematic error.
\end{abstract}

\begin{center}
\rule{8.0cm}{0.2mm}
\end{center}
 
\section{INTRODUCTION}

\begin{center}
\epsfig{figure=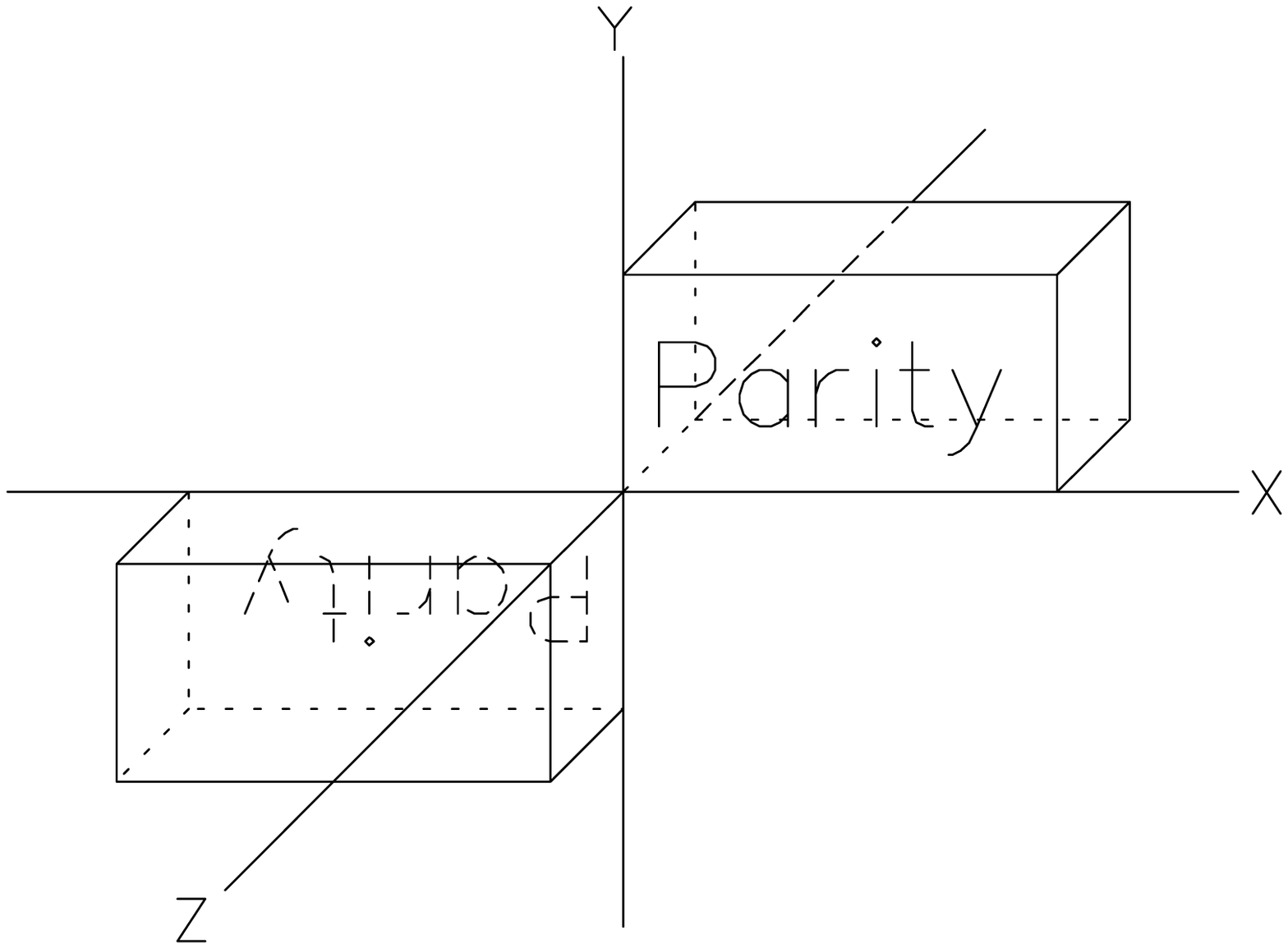, width=12cm}\\[2mm]
\parbox{\linewidth}
{\small \setlength{\baselineskip}{2.6ex}
Fig.~1. \ The parity operation flips all three axes, which is
equivalent to a mirror reflection plus a 180$^0$ rotation. In this
example, it would be a reflection in the $yz$-plane and a rotation
about the $x$-axis. Because we can assume rotational invariance,
parity inversion is often thought of as a mirror reflection.
}
\end{center}

Measuring the effects of the Weak Interaction between two protons in
the presence of the  Strong Interaction, some ten million times
stronger, presents a unique set of difficulties.  The measurement is
only made possible by the fact that the strong force treats
right-handed (spin aligned along the direction of motion like a
right-handed screw) and left-handed particles equally, while the weak
force favours left-handed particles. In fact, of the $W$ and $Z$
bosons which carry the weak force, the $W$s act {\em only} on
left-handed particles, and the $Z$ acts much more strongly on
left-handed particles. In the jargon of the field, the strong
interaction is said to ``conserve parity'' or to be ``invariant under
the parity transformation'', a transformation that reflects each
coordinate through the origin ($\vec{r} \rightarrow \vec{-r}$, see
figure 1). The weak interaction, on the other hand, is not invariant
under such a transformation, and is said to ``violate parity''. In an
experiment now underway at TRIUMF, a beam of protons with their spins
aligned along the direction of motion is  passed through a proton
target (liquid hydrogen), and the fraction scattered when the protons
in the beam have their spins aligned along the beam direction is
compared with the fraction scattered when the spin is opposite to the
beam direction.  Because these two cases differ only in that they are
mirror images of each other, any difference will be an indication of
the ``parity violating'' weak interaction. The difference is expressed
as the parity violating longitudinal analyzing power $A_{z} =
(\sigma^{+} - \sigma^{-}) /(\sigma^{+} + \sigma^{-})$ where
$\sigma^{+}$ and $\sigma^{-}$ are the scattering probabilities for
positive (right-handed) and negative (left-handed) helicity
respectively. The TRIUMF experiment expects to measure $A_z$ with a
statistical uncertainty of $\pm 0.2 \times 10^{-7}$ and a systematic
error of $\pm 0.2 \times 10^{-7}$. The major technical difficulty to
be overcome in such an experiment is to be sure that when the proton
spin is ``flipped'' from right handed to left handed, that the change
in all other beam properties is either zero, or small and measured
accurately enough that its effects can be corrected. The TRIUMF
experiment measures, and corrects for the effects of, beam current,
position, direction, size, and transverse polarization. Most of the
time in such an experiment is spent understanding and controlling all
the sources of systematic error. The current measurement is being made
at 221 MeV, where the theoretical interpretation of the result is
simplified as explained in the next section.

\subsection{Choice Of Energy} 

In describing the scattering of an incident parallel beam of
particles,  it is common in nuclear physics to express the incident
quantum mechanical ``plane wave'' in terms of a sum of ``partial
waves'', each with a specific angular momentum which is conserved in
the scattering.  The effect of a scattering interaction can then be
described in terms of the phase shift it gives to each partial wave.
Figure 2 shows the analyzing power, $A_z$, which we are measuring,
broken down into contributions from the lowest three partial wave
mixings. At low energies, $A_z$ arises almost completely from the
$^{1}S_{0}-^{3}P_{0}$ parity-mixed partial wave\footnote{For
historical reasons, the archaic notation $^{2S+1}\ell_{J}$ is normally
used, where $\ell$ is the orbital angular momentum, with $\ell =
S,P,D,F$... denoting $\ell = 0,1,2,3...$ respectively, $J$ is the
total angular momentum, and $S$ is the total spin. Since parity is
given by $(-1)^{\ell}$, $S$ and $P$ have opposite parity as do $P$ and
$D$.}, while at the energy of the TRIUMF measurement, this
contribution vanishes and $A_z$ arises essentially exclusively from
the $^{3}P_{2}-^{1}D_{2}$  parity-mixed partial wave.

\begin{center}
\epsfig{figure=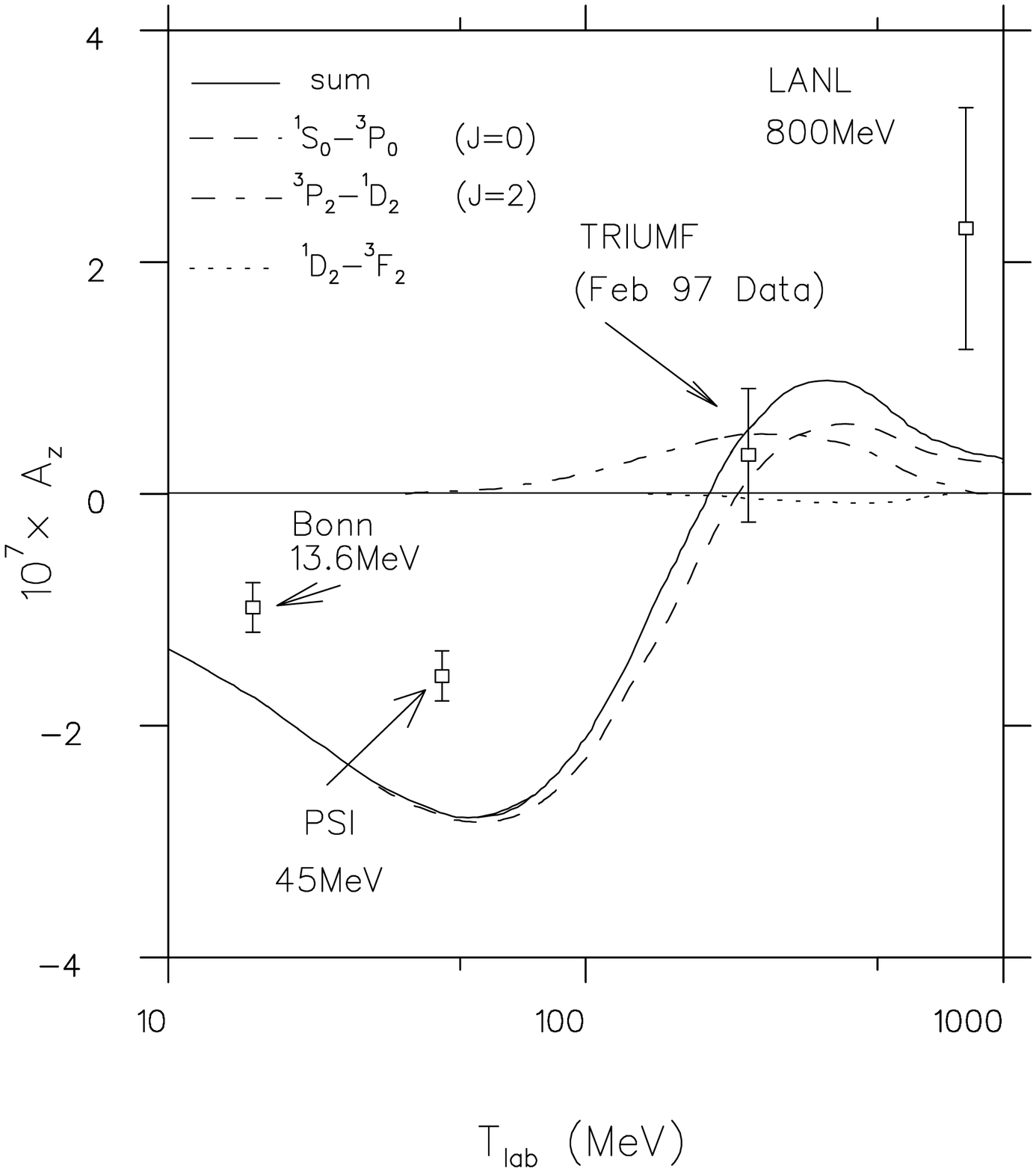, height=8cm, width=10cm}\\[2mm]
\parbox{\linewidth}
{\small \setlength{\baselineskip}{2.6ex}
Fig.~2. \ Contributions to the parity violating longitudinal
analyzing power, $A_z$, as calculated by Driscoll and Miller [1] for
the first three parity mixed partial waves. The energy of the TRIUMF
experiment is chosen so that only the PD wave contributes. Data taken
at Bonn [2], PSI [3], and LANL [4] are also shown. The TRIUMF point is
preliminary, from Hamian's [8] analysis of one data set. The
experiment is designed to obtain a final uncertainty similar to Bonn
and PSI. 
}
\end{center}

Because it is impossible to tell whether a given proton was scattered
by the strong or the weak interaction, the rules of quantum mechanics
dictate that we add the complex ``scattering amplitudes'' rather than
the cross-sections (scattering probabilities). In the expression for
$A_z$, all that remains are the ``interference terms'' involving the
product of strong and weak interactions. The nature of the two
interactions is such that the {\em shape} of the $A_z$ curve is set by
the strong interaction while the {\em sign and absolute scale} is set
by the weak interaction.  Since the strong proton-proton interaction
is already very well measured, it is possible, without knowing the
weak interaction, to accurately determine the energy at which the $SP$
contribution crosses zero.

Just as the electromagnetic force between two electrons is interpreted
as due to the exchange of photons between them, the strong force
between two protons can be described by the exchange of mesons between
them. The analyzing powers in figure 2 were calculated by Driscoll and
Miller\cite{dm} using such a meson exchange model.  In this
calculation, the $SP$ contribution arises roughly equally from $\rho$
and $\omega$ meson exchange, whereas the $PD$ contribution arises
almost exclusively from $\rho$-meson exchange.  By measuring at an
energy where the $SP$ contribution integrates to zero over the
acceptance of our apparatus, the TRIUMF experiment will be able to
determine the value of the weak $\rho$ meson-nucleon-nucleon coupling
constant $h_\rho$, a number that is now known only very poorly. Figure
3 shows the $PD$ contribution for a range of $h_\rho$ considered
``reasonable'' by Desplanques, Donoghue, and Holstein \cite{ddh}. The
TRIUMF measurement will constrain this range significantly.

Also significant is the fact that at 221 MeV, the
$^{3}P_{2}-^{1}D_{2}$ wave corresponds to an interaction near the
surface of the proton.  This probes an intermediate region between the
long range interaction which can be considered strictly meson exchange
and the very short range part where the interaction takes place
between the quarks of the proton. The extent to which quark degrees of
freedom are important here is an interesting question. For example, 
taking explicit account of quark degrees of freedom, Grach and
Shmatikov \cite{gs} calculate $A_{z} = 2.4 \times 10^{-7}$ at 230 MeV
compared to the $A_{z} = 0.6 \times 10^{-7}$ from the meson exchange
calculation of Driscoll and Miller \cite{dm} referred to earlier.
Adding an intermediate $\Delta$ to the meson exchange calculation,
Iqbal and Niskanen \cite{in} calculate $A_{z} = 1.1 \times 10^{-7}$ at
the same energy.  Clearly a good measurement is needed.

\begin{center}
\epsfig{figure=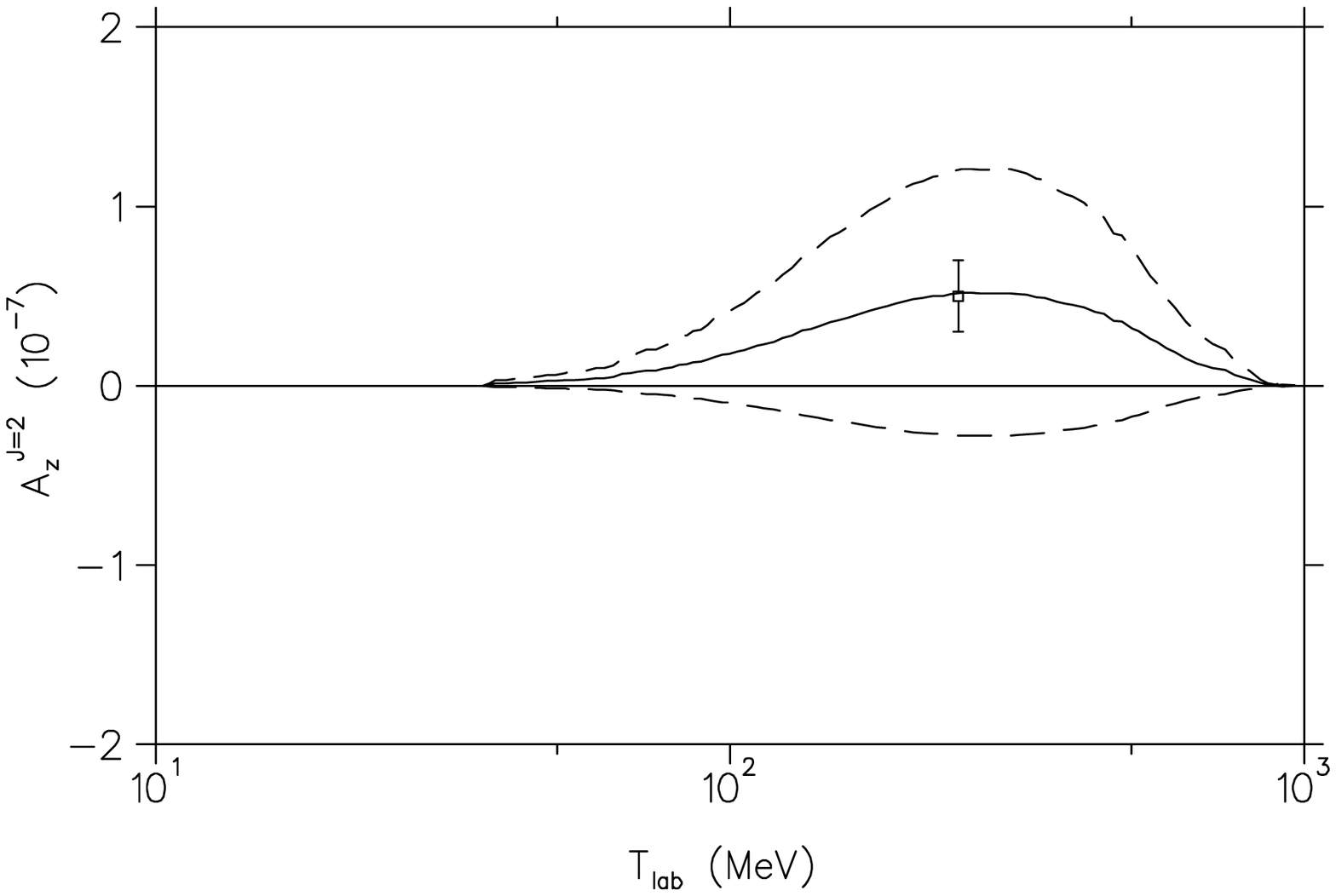, height=6cm}\\[2mm]
\parbox{\linewidth}
{\small \setlength{\baselineskip}{2.6ex}
Fig.~3. \ Contribution to $A_z$ from the parity mixed partial wave
(PD) to which the TRIUMF experiment is sensitive. The solid curve
shows the result of Driscoll and Miller's [1] calculation using the
DDH [5] ``best guess'' value for  $h_\rho$, and the dotted curves show
what would be expected if $h_\rho$ varied over the DDH ``reasonable
range''.  The fictitious data point shows the error bar expected from
the TRIUMF measurement and illustrates how such a measurement will
significantly constrain the range of $h_\rho$.
}
\end{center}

\section{THE EXPERIMENT}

\begin{center}
\epsfig{figure=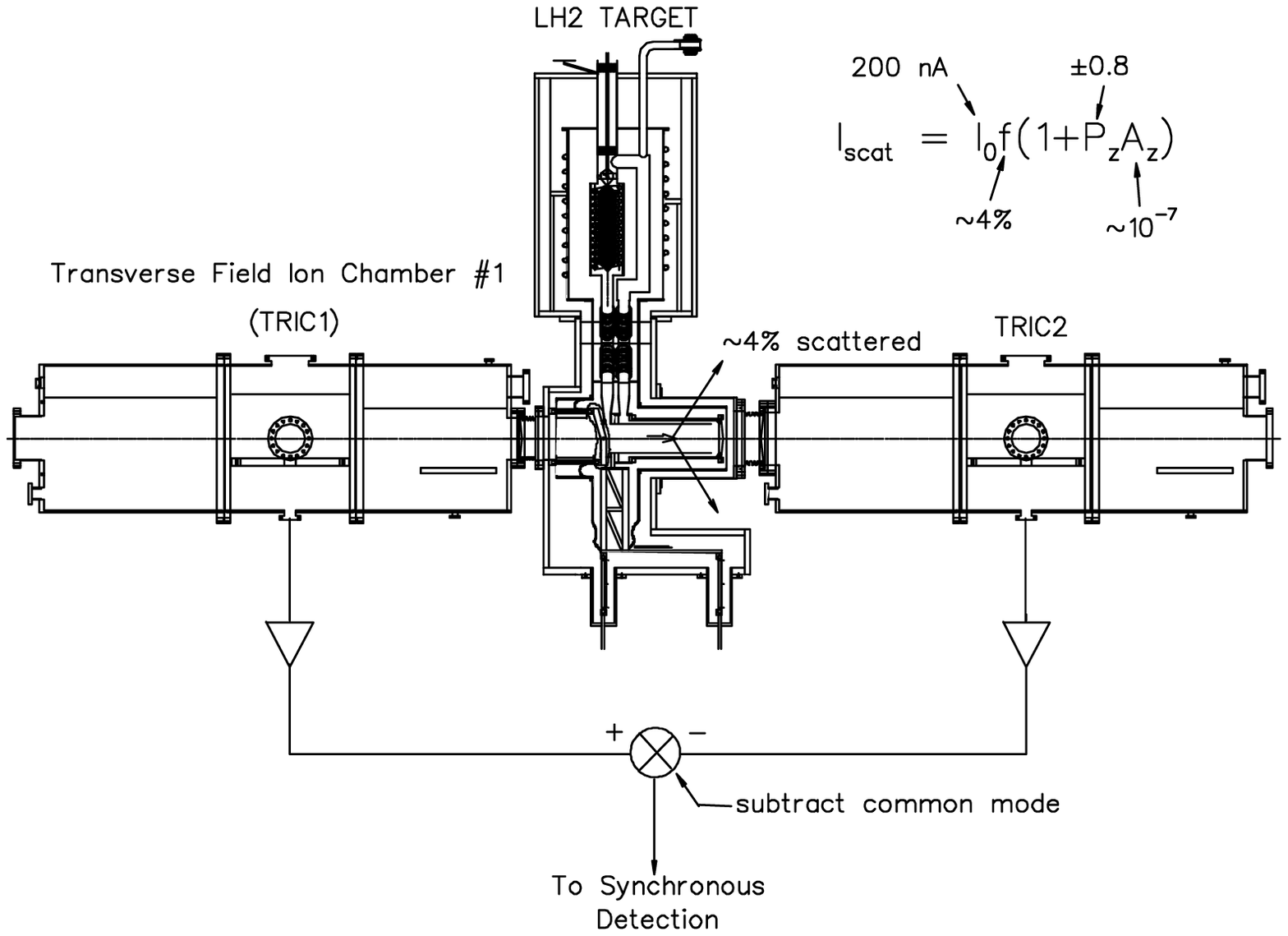, width=13cm}\\[2mm]
\parbox{\linewidth}
{\small \setlength{\baselineskip}{2.6ex}
Fig.~4. \ Heart of the experiment, showing the target and main
detectors. TRIC1 and TRIC2 are ion chambers containing hydrogen gas at
approximately $1/5$ atmosphere pressure.  The target is 40 cm of
liquid hydrogen. 
}
\end{center}

Figure 4 shows the principle of the experiment. A 200 nA beam of
protons with energy 221 MeV passes through the first detector
(Transverse Ion Chamber 1, or TRIC1), then 40 cm of liquid hydrogen,
and finally through the second detector (TRIC2).  The first detector
measures the beam current before the target, the second detector
measures the beam current after the target, and a precision subtractor
takes the difference between the two signals. The detector and
electronic gains are adjusted to make the output of the subtractor
zero with unpolarized beam\footnote{For those interested in technical
details, I point out that we are interested in the AC part of the
signals, and that the fine gain adjustment is made for best
common-mode rejection at the spin flip frequency. This reduces both
shot noise and the sensitivity to helicity correlated changes in beam
current.}. If there is a parity violating analyzing power for $pp$
scattering at this energy, then the scattered fraction ($\sim$4\%)
will be slightly more or slightly less than the unpolarized case
depending on the helicity of the beam. A signal synchronized with the
helicity and whose magnitude is proportional to $A_z$, will appear at
the output of the subtractor. This signal is very small. For example,
if $A_z = 0.6 \times 10^{-7}$ and the longitudinal polarization is
80\%, we must measure a difference signal only $2 \times 10^{-9}$ of
the TRIC signal, a feat comparable to measuring the thickness of a
piece of paper in a 40 km measurement. 

Upstream of TRIC1 are Intensity Profile Monitors and Polarization
Profile Monitors (IPMs and PPMs in figure 7) which measure the
distribution of intensity and transverse polarization across the beam.

\subsection{Data Acquisition}

To recover such a small signal, we use synchronous detection, the same
principle used by lock-in amplifiers.  The data are binned according
to spin state and differences which are uncorrelated with spin state
average out with time, whereas helicity correlated differences do not.

The data are taken in $\frac{1}{5}$ second (200 ms) cycles, each cycle
consisting of eight $\frac{1}{40}$ second (25 ms) spin states arranged
in the pattern $(+--+-++-)$ or its complement. The cycles are further
arranged in a ``super-cycle'' with the starting spin state of each
cycle following the same eight state pattern.  The starting spin state
of each supercycle is chosen at random. This data taking pattern
cancels both linear and quadratic drifts. 

Each 25 ms spin state is divided into polarization measuring and
asymmetry measuring intervals as shown in figure 5. The main
integration period is set for exactly $\frac{1}{60}$ second to reject
60 Hz line frequency noise and all its harmonics.

\begin{center}
\epsfig{figure=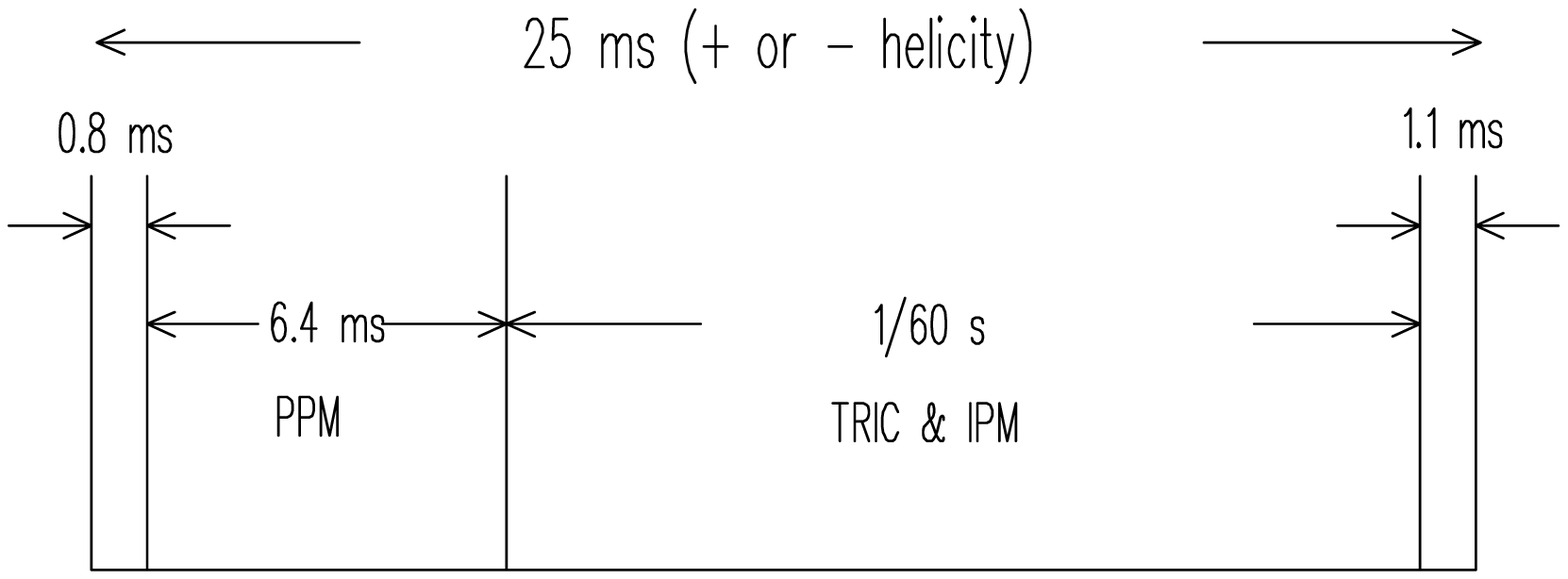, width=10cm}\\[2mm]
\parbox{\linewidth}
{\small \setlength{\baselineskip}{2.6ex}
Fig.~5. \ Intervals during each spin state. First the  PPM
(Polarization Profile Monitor) scans a $CH_2$ blade through the beam
to measure the polarization distribution across the beam, then the
beam currents and profiles are measured by the TRICs (Transverse Ion
Chambers) and IPMs (Intensity Profile Monitors).
}
\end{center}

\subsection{Sources Of Error}

Sources of error can be divided into those which are related to the
beam helicity and those which are not. Beam property changes which are
synchronized with spin flip (``coherent'' or ``helicity correlated''
changes) can, and usually do, produce a false signal of parity
violation. Things which are not helicity correlated, such as detector
noise and random variation in beam properties, do not bias the result,
but only increase the run time required to reach a given precision. 

\subsubsection{\underline{\normalfont RANDOM CHANGES}}

The ultimate limit to the statistical precision of the experiment is
that determined by the counting statistics of the scattered protons, a
limit which could be reached if we were able to count individual 
scattered protons. Then $A_z$ could be measured to $\pm 0.2 \times
10^{-7}$ in 20 hours. In practice, the count rate of the scattered
protons is too high ($\sim$50 GHz) for direct counting and the
experiment measures currents instead. With our existing detector
configuration, the running time for a $\pm 0.2 \times 10^{-7}$
precision rises to approximately 300 hours, largely because of
detector noise\footnote{We could match this with a counting experiment
by reducing the beam current to count at $\sim$1.5 GHz. This rate is
still impractically high and it is easier to use the higher beam
current and put up with detector noise.}.

In addition to detector noise, random variations in beam properties
such as intensity or position contribute to noise in the signal and
increase the required run time.

\subsubsection{\underline{\normalfont HELICITY CORRELATED CHANGES}}

Helicity correlated changes in beam properties can produce a false
$A_z$ signal. This false signal is dealt with in different ways:
\begin{itemize}
\item Careful design and operation of the TRIUMF polarized ion source
and cyclotron makes it possible to change the spin direction with the
absolute minimum of effect on other beam properties.
\item The parity equipment and operating conditions are carefully
chosen to minimize the sensitivity to helicity correlated changes.
\item The beam properties are accurately measured during data taking
so the actual helicity correlated changes are known for each data set.
\item Calibration runs determine the sensitivity to helicity
correlated modulations so that corrections can be made. 
\end{itemize}
Transverse polarization, beam intensity, position, and size are all
measured and the resultant false effects are corrected for.

\begin{center}
\epsfig{figure=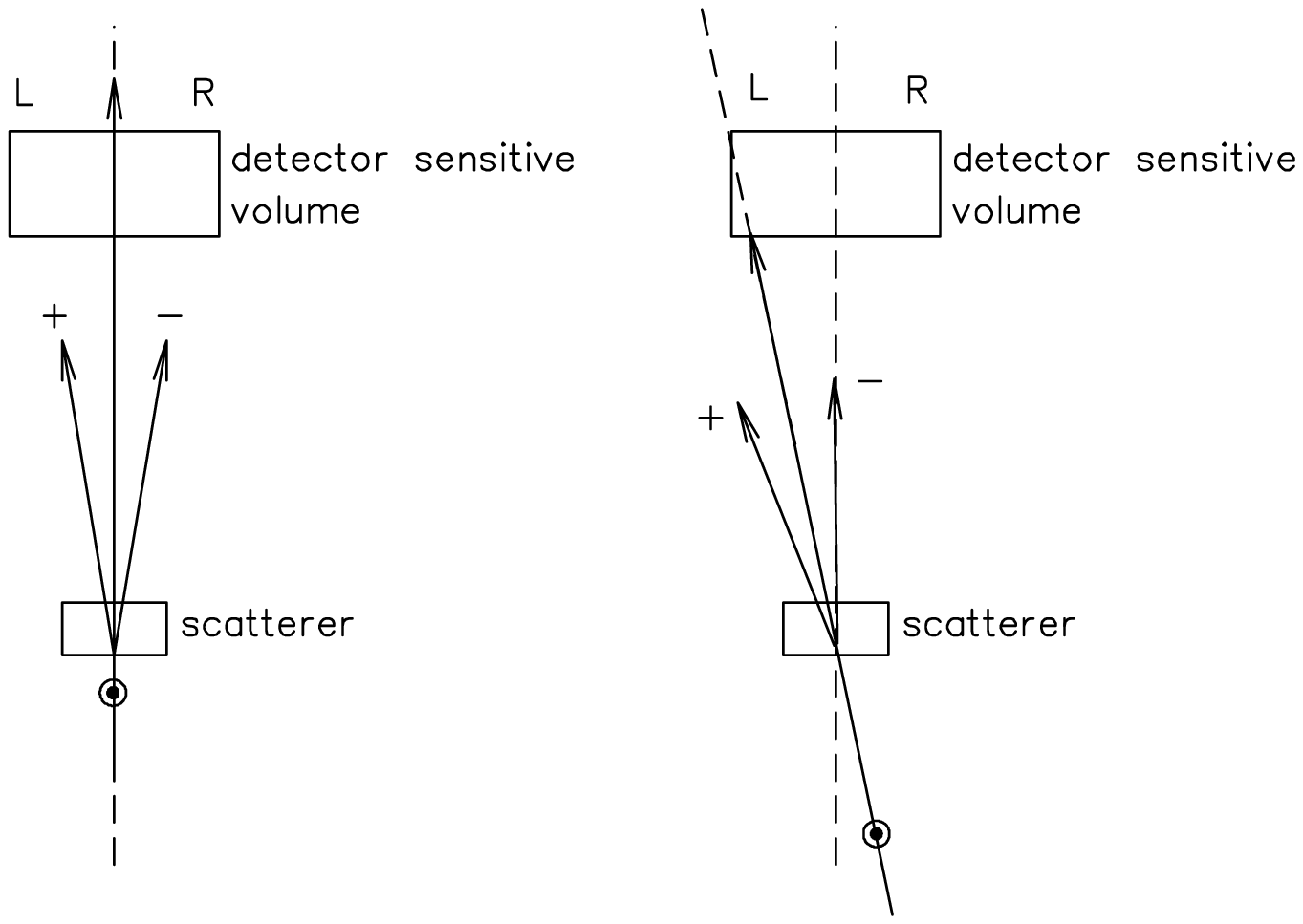, width=10cm}\\[2mm]
\parbox{\linewidth}
{\small \setlength{\baselineskip}{2.6ex}
Fig.~6. \ The effect of transverse components of polarization.  The
two diagrams use a pencil beam with vertical polarization to
illustrate the mechanism. In both cases, because the strong
interaction analyzing power is positive in this example, the scatterer
causes more beam to be scattered to the left in the positive spin
state and more to the right in the negative spin state.  In the
left-hand diagram, since the beam passes through the center of the
detector, there is no difference in the detector response for the two
spin states.  In the right-hand diagram, however, the beam is left of
center at the detector, so the signal is larger in the negative spin
state, causing a false signal of parity violation. The false signal is
proportional to the size of the transverse component multiplied by the
distance off center -- a quantity called the {\em first moment} of
transverse polarization.  A real beam of finite extent is made of a
bundle of such pencil beams, and can have a first moment of transverse
polarization even if its average transverse polarization is zero. For
example, spin could be up on one side of the beam and down on the
other.
}
\end{center}

\paragraph{\underline{\normalfont Transverse Polarization}}

If the proton spin is not perfectly longitudinal, the small transverse
component will reverse with helicity.  As explained in the caption to
figure 6, the false signal is proportional to the product of the
transverse component and the distance that the beam is off center.
This product is referred to as the {\em first moment} of transverse
polarization.  First moments of transverse polarization are the most
difficult property for us to measure and are the main source of
systematic error.

To determine first moment sensitivities, test runs are made with pure
vertical and pure horizontal polarization.  By scanning the vertically
polarized beam horizontally and the horizontally polarized beam
vertically, we are able to determine our sensitivities to the first
moments $xP_y$ and $yP_x$, and to find the beam position at which
there is no sensitivity to average transverse polarization. This axis,
where transverse polarization components cause no false effect, is
called the {\em polarization neutral axis}.

In analyzing the effect of first moments, we separate {\em extrinsic}
first moments caused by a beam whose centroid is displaced from the
neutral axis and which has some average transverse polarization, and
{\em intrinsic} first moments which do not depend on the position of
the beam centroid but rather arise from the distribution of transverse
components within the beam. By using fast ferrite-cored steering
magnets servoed to the intensity profile monitors (IPMs in figure 7),
we are able to hold the beam on the polarization neutral axis and
virtually eliminate corrections for {\em extrinsic} first moments. 
{\em Intrinsic} first moments, on the other hand are independent of
beam centroid position, arise in the cyclotron and beamline, and are
very hard to control. Corrections for intrinsic first moments are
typically the only corrections which are not consistent with zero when
averaged over a one month running period.

By measuring the distribution of transverse components of polarization
at the two Polarization Profile Monitors (PPMs in figure 7) we are
able to use our measured first moment sensitivities to correct the
data for the effects of first moments.  The PPMs are able to determine
first moments to $\pm 5 \mu m$ in a one hour run. For a one month data
taking period, the final correction to the raw $A_z$ is substantial,
of the order $10^{-7}$, and uncertainty in the correction increases
the uncertainty in the final $A_z$. For example, in the
Hamian's\cite{aah} analysis of the February, 1997 data set, the
statistical uncertainty of $\pm 0.58 \times 10^{-7}$ was increased to
$\pm 0.65 \times 10^{-7}$, primarily by uncertainty in the first
moment corrections.

\begin{center}
\epsfig{figure=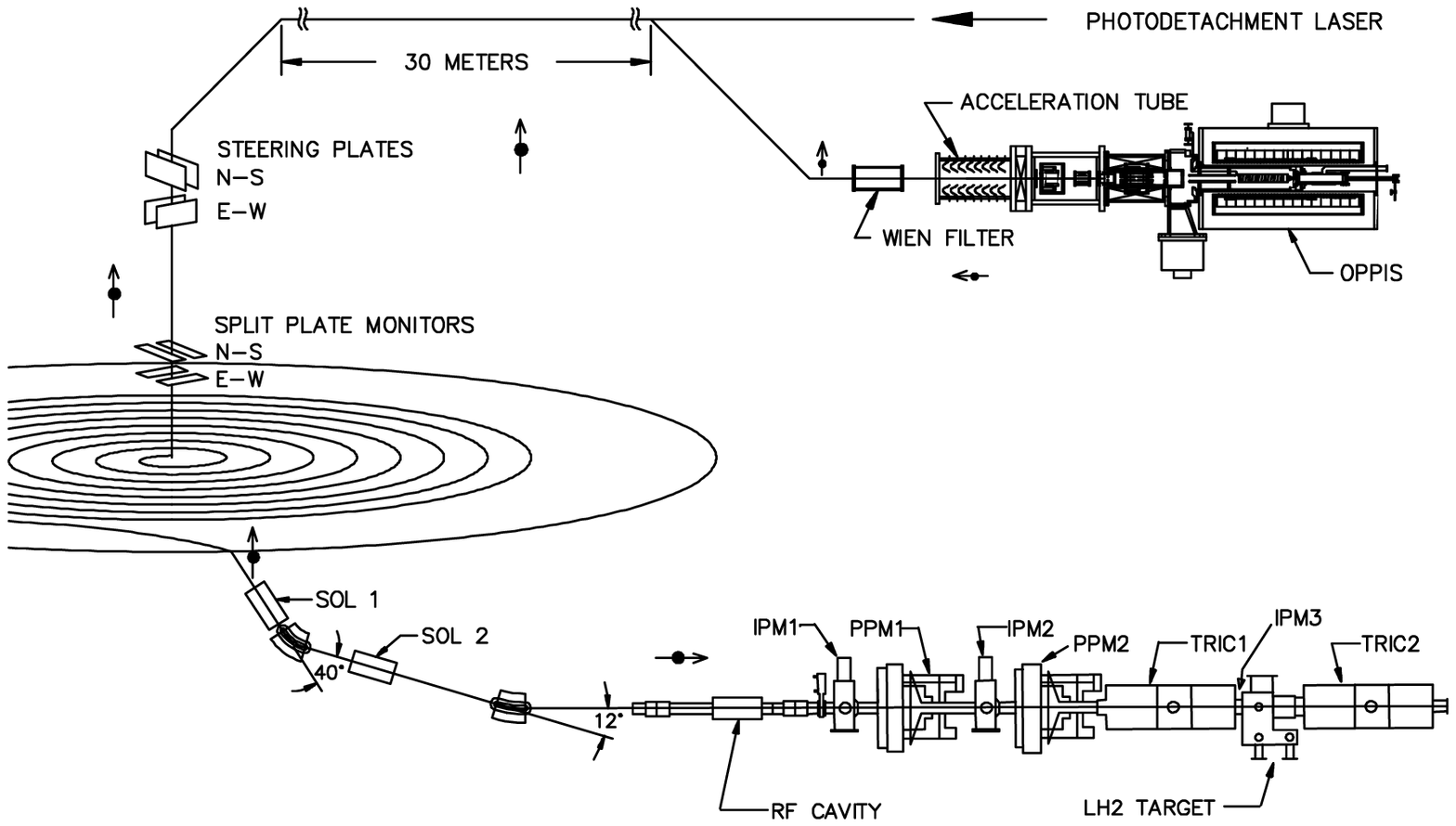, width=\linewidth}\\[2mm]
\parbox{\linewidth}
{\small \setlength{\baselineskip}{2.6ex}
Fig.~7. \ Schematic of the whole Parity Experiment. Every item, from
the ion source (OPPIS) to the end of the beamline, must be optimized
to reduce systematic errors.
}
\end{center}

\paragraph{\underline{\normalfont Beam Intensity}}

If the gains are properly set, the beam currents before and after the
target subtract perfectly and we are blind to beam current changes. In
practice this ``common mode rejection'' is good\footnote{Of the order
80 to 90 db.}, but is not perfect, and it drifts slightly with time.

During normal operation, the coherent intensity modulation which
accompanies spin flip ranges from approximately $2 \times 10^{-5}$ to
$8 \times 10^{-5}$ of the 200 nA beam current. To determine what
effect this has on the measured $A_z$, 1.6 seconds out of every 16 are
devoted to running with the ion source unpolarized and the beam
intensity artificially modulated by a laser (``photodetachment laser''
in figure 7) which co-propagates with about 30 m of the $H^-$ beam
prior to injection into the cyclotron.  When this laser is on,
electrons are removed from some of the $H^-$ ions and these are then
not accelerated. This gives us data with large ($\sim 0.2$\%) pure
intensity modulation interleaved with the primary data. The false
$A_z$ produced by the photodetachment laser is then scaled using the
measured value of intensity modulation recorded during the real data,
and a correction can be made. 

\paragraph{\underline{\normalfont Beam Position Modulation}}

The beam profiles are measured during each spin state using two $x-y$
Intensity Profile Monitors (IPM1 and IPM2 on figure 7). From this the
beam centroid, size, and skewness are calculated and the mean and
helicity correlated values are extracted.  The false $A_z$ arising
from helicity correlated motion is proportional to the size of the
motion and the distance the beam is off the ``neutral axis'' of the
experiment. To calibrate the sensitivity to coherent position
modulation, a series of test runs are made in which fast ferrite-cored
steering magnets are used to intentionally introduce a series of large
coherent position changes. Enough different beam positions and
modulations are recorded to describe our sensitivity to position
motion. During real data taking we find that the actual beam position
modulation with spin flip is at the limit of our detection ability
($0.1 \mu m$ in a one-hour run) and overall corrections for position
modulation are insignificant.

\paragraph{\underline{\normalfont Beam Size Modulation}}

The same Intensity Profile Monitors which determine the beam position
for each spin state also determine the beam size. If the beam size
changes on spin flip, more or less of the beam tails will be clipped
by the finite detector aperture and a false signal will appear. To
calibrate our sensitivity to size modulation, large, intentional size
modulation is introduced using two fast ferrite-cored quadrupole
magnets. We find that our apparatus is very sensitive to size
modulation. Fortunately the actual coherent size modulation with spin
flip is very small ($< 0.1 \mu m$ on a $5 mm$ beam) and its effect on
our final $A_z$ averages to near zero.

\paragraph{\underline{\normalfont Energy Modulation}}

The gain of the hydrogen-filled ion chambers used as our main
detectors is given by the number of electron-ion pairs produced by a
proton going through the active region. As the beam energy is lowered,
the energy loss per unit length ($\frac{dE}{dx}$) goes up and the gain
increases. Because the proton beam loses approximately 27 MeV in the
liquid hydrogen target, the gas gain in the downstream chamber rises
more rapidly with a drop in beam energy than does the gain in the
upstream chamber.  The result is that a coherent beam energy
modulation produces a false $A_z$ signal.  In fact, the experiment is
very sensitive to coherent energy modulation with
$\partial{A_z}/\partial{E} = 2.8 \times 10^{-8}$ per eV. 

It is normally possible to keep energy modulation at the ion source at
the few {\em milli}-eV level, and it is known that the injection
system and cyclotron will multiply this by a factor of about 100, so
we believe that coherent energy modulation does not bias our $A_z$
significantly.  Nevertheless, the amplification factor can vary quite
a bit and we are not able to measure the coherent energy modulation
directly to the required precision during running, so another approach
is adopted to minimize such effects.

It is possible to tune our beamline so that spin-up from the cyclotron
becomes either positive or negative helicity at the parity apparatus.
If the spin-up state always has, say, a slightly higher energy than
the negative state, then when we switch the beamline helicity,
asymmetry arising from true $A_z$ will reverse, but that from energy
modulation will not.  By averaging the {\em apparent} $A_z$ from the
two different beamline helicities, effects of energy modulation should
cancel out.

\subsection{State Of The Data}

In experiments such as this, most of the time is spent doing
development and control measurements to understand and minimize
sources of systematic error. Following some years of developing
detectors and diagnostic equipment, the experiment was finally mounted
on a new beamline 4A2 at TRIUMF in late 1994.  1995 and much of 1996
were spent improving the performance of the TRIUMF cyclotron and ion
source, and refining our systematic error controls.  A major data
taking run was made in February and March of 1997.  This data has been
analyzed and a preliminary result reported by Hamian\cite{aah}. She
obtained an overall uncertainty of $\pm 0.65 \times 10^{-7}$. Another
major run from July and August of 1998 is now being analyzed. With the
addition of a run planned for the summer of 1999, we should have
enough data to determine $A_z$ with an uncertainty of $\pm 0.3 \times
10^{-7}$, adequate to significantly improve our knowledge of the weak
meson-nucleon-nucleon coupling constant, $h_\rho$.

\section{CONCLUSION}

Proton-proton parity violation experiments are technically demanding,
but provide a unique window on the interplay of strong and weak
interactions.  The TRIUMF 221 MeV measurement probes a very
interesting region near the surface of the proton, between the quarks
inside the proton and the pion exchange of more distant interactions.
In terms of meson exchange, it is sensitive only to the $\rho$-meson
exchange. The experiment is expected to finish data taking in 1999 and
a final result should be available in 2000.

\vspace{0.2cm}
\bibliographystyle{unsrt}

\end{document}